\documentclass[a4paper,fleqn,useAMS,usenatbib]{mnras}
\usepackage[T1]{fontenc}
\usepackage{ae,aecompl}

\usepackage{aas_macros}
\usepackage{graphicx}
\usepackage{amssymb}
\usepackage{amsmath}
\usepackage{widetext}
\def\psad{$P^2SAD$}
\usepackage{color}

\title[Universal clustering of dark matter]{Universal clustering of dark matter in phase space }
\author[Jes\'us Zavala, Niayesh Afshordi]{\parbox{18cm}{Jes\'us
    Zavala$^{1}$\thanks{Marie Curie Fellow; e-mail: jzavala@dark-cosmology.dk} and Niayesh Afshordi$^{2,3}$\vspace{0.3cm}}\\ 
$^{1}$Dark Cosmology Centre, Niels Bohr Institute, University of Copenhagen, Juliane Maries Vej 30, 2100 Copenhagen, Denmark\\
$^{2}$Perimeter Institute for Theoretical Physics, 31 Caroline St. N., Waterloo, ON, N2L 2Y5, Canada\\
$^{3}$Department of Physics and Astronomy, University of Waterloo, Waterloo, Ontario, N2L 3G1, Canada\\}

\date{Accepted XXX. Received YYY; in original form ZZZ}

\pubyear{2015}

\begin{document}
\label{firstpage}
\pagerange{\pageref{firstpage}--\pageref{lastpage}}
\maketitle

\begin{abstract}
We have recently introduced a novel statistical measure of dark matter clustering in phase space, the 
particle phase space average density (\psad). In a two-paper
series, we studied the structure of \psad~in the Milky-Way-size Aquarius haloes, constructed a physically motivated model to describe it, and illustrated its potential
as a powerful tool to predict signals sensitive to the {\it nanostructure} of dark matter haloes. In this work, we report
a remarkable universality of the clustering of dark matter in phase space as measured by \psad~within the subhaloes of host
haloes across different environments covering a range from dwarf-size to cluster-size haloes ($10^{10}-10^{15}$~M$_\odot$). Simulations show that the universality of \psad~holds
for more than 7 orders of magnitude, over a 2D phase space, covering over 3 orders of magnitude in distance/velocity, with a simple functional form that can be described by our model. 
Invoking the universality of \psad, we can accurately predict the non-linear power spectrum of dark matter at small scales all the way down to the decoupling mass limit of cold dark matter particles. As an application, we compute the subhalo boost to the annihilation of dark matter in a wide range of host halo masses. 
\end{abstract}


\section{Introduction}

The Cold Dark Matter (CDM) paradigm of structure formation predicts the clustering of dark matter (DM) into gravitationally bound haloes in a very large range of scales, from 
the decoupling of CDM particles (e.g., $10^{-11}-10^{-3}$~M$_\odot$, \citealt{Bringmann_09}, depending on the model) to the 
massive $10^{15}$~M$_\odot$ clusters virialized recently. Because in CDM most of the DM today is predicted to be inside haloes, accurately following the evolution of the
DM phase space distribution within these highly non-linear structures is a challenging task. The tremendous progress of numerical $N-$body simulations have made it 
possible to cover the dynamical range paramount to galaxy formation, from large ($\sim$Mpc) to subgalactic ($\sim$100~pc) scales, but is not yet feasible to explore the DM clustering at even lower scales, which we refer to as the {\it nanostructure}
of DM haloes. This unresolved regime is however of prime
importance in experiments searching for non-gravitational DM signatures
that use theoretical predictions, which
in many cases are quite sensitive to the {\it nanostructure} of DM haloes. 
For instance, the DM self-annihilation rate over an entire halo is dominated by
events occurring in the plethora of its unresolved subhaloes, while the scattering rate in direct detection experiments is sensitive to the local 
DM phase space distribution.    

\begin{table*}
\centering
\begin{tabular}{cccccccc}
\hline
Simulation  & $M_{200}$[M$_\odot$] & $r_{200}$[Mpc]& $m_p$[M$_\odot$] & $\epsilon$[kpc] & $M_{\rm sub}$[M$_\odot$] & $N_{\rm sub}$[$\times10^5$] & {\rm Reference}\\
\hline
\hline
Dwarf-size    &   $1.3\times10^{10}$  & $0.047$ & $1.2\times10^3$    & $0.034$  & $3.2\times10^{8}$ & 2.69 & \citet{Vogelsberger_14} \\  
Milky-Way-size    &   $1.8\times10^{12}$  & $0.246$ &  $1.4\times10^{4}$ & $0.066$ & $1.3\times10^{11}$ & 96.14 & \citet{Springel_08} \\  
Group-size   &   $8.7\times10^{13}$  & 0.939 & $1.5\times10^7$    & $1.6$ & $6.8\times10^{12}$ & 4.55 & {\rm this work}\\  
Cluster-size    &  $2.9\times10^{15}$   & 3.022 & $2.4\times10^8$    & $4.0$ & $1.5\times10^{14}$ & 6.29 & {\rm this work}\\  
\hline
\end{tabular}
\caption{Summary of some of the properties of the simulations we analyze (at $z=0$): the virial mass ($M_{200}$) and radius ($r_{200}$), the particle mass ($m_p$), the Plummer-equivalent maximum physical softening length ($\epsilon$), the total mass ($M_{\rm sub}$), the total number of particles ($N_{\rm sub}$) in subhaloes resolved with more than 20 particles, and the reference to each simulation. Except for the Group-size case, all simulations have a lower resolution version (by a factor of $2$ in softening), that we use to test convergence in \psad.}
\label{table_sims} 
\end{table*}

Even if there are no new DM interactions with visible particles, it might still be possible to detect the DM clustering at small scales through the gravitational influence of DM on astrophysical
sources. Although haloes with a mass below the limit for atomic cooling are expected to be devoid of cold gas and stars, their presence can still be made evident through the gravitational lensing
they produce on background sources. For instance, subhaloes close to the atomic cooling limit ($\sim10^6-10^8$M$_\odot$) might be responsible for the lensing flux-ratio anomalies observed in quasars \citep[e.g.][]{Xu_15}, while even smaller subhaloes could be probed by time-varying lensing effects as they move along the line of sight of cosmological sources such as pulsars and quasars \citep{Baghram_11,Rahvar_14}. 

The spatial DM clustering at subhalo scales is known to have nearly universal properties within the range of scales resolved in simulations: a smooth radial distribution described by a spherically averaged two-parameter NFW profile \citep{NFW_97}, and a hierarchy of subclumps (well described by NFW profile truncated at the tidal radius) with an abundance that is a power-law in mass, the subhalo mass function \citep[e.g.][]{Springel_08}. 
Although the properties of the hierarchy of subhaloes are complicated by the time-varying effects of tidal disruption, (extrapolating) these observed universalities are the basis of most predictions of the DM clustering at small, unresolved scales.

A complimentary picture emerges by considering the DM clustering in phase space, which becomes particularly relevant in potential signals of new DM interactions that are sensitive to the DM velocity distribution. In a series of papers \citep[][henceforth Papers I and II]{Zavala_14a, Zavala_14b}, we introduced the two-dimensional particle phase space average density (\psad), a coarse-grained phase space density, which is a novel
measure of DM clustering at small scales. In Paper I, we found signs of a (near-)universality in the structure of \psad~over assembly history and redshift for Milky-Way-size haloes, while in Paper II we presented a physical model of \psad~based on the stable clustering hypothesis in phase space \citep{Davis_77, Afshordi_10}, the spherical collapse model, and tidal disruption of subhaloes. We then showed how this model can be used to predict DM annihilation signals. 

In this work, we investigate further the structure of \psad~and find that is remarkably universal for DM inside subhaloes across a wide range of host halo masses. It has a functional form that can be described by a simple parametric formula with a structure that is motivated (modelled) by our physical prescription. Such universality in phase space makes it a powerful tool to describe the DM clustering at small unresolved scales, while its simplicity makes it useful to predict several potentially observable DM signals.

\section{Particle Phase Space Average Density (\psad) in subhaloes}\label{2PCF}

We follow the same notation as in Paper I to define \psad~$\equiv\Xi(\Delta x, \Delta v)_{{\cal V}_6}$ as the mass-weighted average (over a volume ${\cal V}_6$ in phase space) 
of the coarse-grained phase space DM density, on spheres of radius $\Delta x$ and $\Delta v$, in position and velocity 
spaces, respectively:
\begin{eqnarray}\label{coarse}
\Xi(\Delta x, \Delta v)_{{\cal V}_6}&\equiv&\frac{\int_{{\cal V}_6}d^3{\bf x}d^3{\bf v} f({\bf x},{\bf v})f({\bf x}+{\bf\Delta x},{\bf v}+{\bf \Delta v})}
{\int_{{\cal V}_6} d^3{\bf x}d^3{\bf v} f({\bf x},{\bf v})}\nonumber\\
&\equiv& \frac{\left<f({\bf x},{\bf v})f({\bf x}+{\bf\Delta x},{\bf v}+{\bf \Delta v})\right>_{{\cal V}_6}}{\left<f\right>_{{\cal V}_6}}
\end{eqnarray}
where $f({\bf x},{\bf v})$ is the phase space distribution function at the phase space coordinates $\bf x$ and $\bf v$, and $\left<f\right>_{{\cal V}_6}$ is
the average phase space density in the volume ${\cal V}_6$:
\begin{equation}
	\left<f\right>_{{\cal V}_6}=\frac{{\int_{{\cal V}_6} d^3{\bf x}d^3{\bf v} f({\bf x},{\bf v})}}{{\cal V}_6}\equiv\frac{M_{{\cal V}_6}}{{\cal V}_6}
\end{equation}

In a simulation, the dark matter density field is represented by a discrete set of $N$ particles, each with a mass $m_p$. In this representation, 
Eq.~(\ref{coarse}) is estimated as:
\begin{equation}\label{2pcf_eq_sim}
\Xi(\Delta x, \Delta v)_{\rm sim}=\frac{m_p \langle N_p(\Delta x, \Delta v)\rangle_{{\cal V}_6}}{V_6(\Delta x,\Delta v)},
\end{equation}
where $\langle N_p\rangle$ is the average number of simulation particles (over ${\cal V}_6$) within shells of thickness $(\delta x,\delta v)$ at a radius 
$\Delta x$ and $\Delta v$ in phase space centred on each of the particles, and 
$V_6(\Delta x,\Delta v)$ is the phase space volume of a given shell (see Fig. 1 of Paper I for a diagram illustrating how \psad~is estimated in a simulation).

In Papers I and II, we studied \psad~averaged over all particles within the virial radius of each of the Aquarius haloes \citep{Springel_08}. 
We define the virial radius as the radius enclosing a sphere with mean density 200 times the critical value ($r_{200}$). Here, we focus instead only on the 
particles inside self-bound subhaloes within the
virial radius of the host, and extend the analysis to different host halo masses, covering a range from dwarf-size  to cluster-size haloes ($10^{10}-10^{15}$~M$_\odot$, see Table~\ref{table_sims} for a summary of the simulations we use).

\begin{figure}
\center{
\includegraphics[height=8.5cm,width=8.5cm]{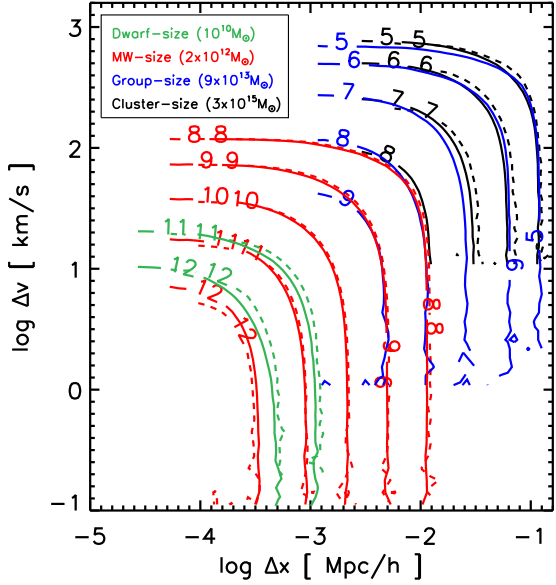} 
}
\caption{Contours of the logarithm of the particle phase space average density (\psad) averaged over all particles inside the subhaloes 
of four haloes of different sizes at $z=0$: cluster-size (black), group-size (blue), Milky-Way-size (red) and dwarf-size (green).
Two resolution levels are shown: high (solid) and low (dashed); see Table~\ref{table_sims} for a summary of the simulations. The average phase-space
DM clustering in subhaloes is remarkably universal across haloes of different masses and environments.} 
\label{fig_p2sad_subs} 
\end{figure}

We find that \psad~is remarkably universal across all halo masses (see Fig.~\ref{fig_p2sad_subs}). In Paper I, we had already reported some level of universality for the limited mass
range of the Aquarius haloes ($\sim8\times10^{11}-2\times10^{12}$M$_\odot$), and for \psad~averaged over all particles within $r_{200}$. Here we show that the universal character of \psad~{\it within subhaloes} extends across the more than 5 orders of magnitude in halo masses that we have explored. Notice the large range of scales
($3$ orders of magnitude in both, distance and velocity separations; 18 orders of magnitude in the 6D phase space volume) where the universality of \psad~holds, while varying by $7$ orders in magnitude.

Subhaloes with a characteristic physical scale and velocity contribute more prominently to a fixed value of \psad, i.e., each contour in Fig.~\ref{fig_p2sad_subs} is mainly representative of subhaloes of similar size. The maximum scale that subhaloes can have is limited by the size of their host (governed by the tidal stripping in the parent halo), and thus, above this maximum scale, subhaloes of a given host cannot contribute to low values of \psad, which is why the subhaloes in the dwarf (Milky-Way) halo only contribute fully up to log(\psad)$\sim11$(8). The resolution of each simulation on the other hand, sets a minimum subhalo mass and thus a maximum value of \psad~that can be trusted. These two scales define the innermost and outermost contours that, for each halo, satisfy the universality of \psad. In Fig.~\ref{fig_p2sad_subs}, we have only plotted the countours within these boundaries.

By taking only the particles within subhaloes, convergence in $P^2SAD$ is harder to achieve due to poorer sampling within the volume ${\cal V}_6$, than in the case when
all particles within the virial radius are used. This is particularly relevant at small scales in phase-space (log(\psad)$>11$) where
the subhaloes that dominate \psad~are sampled with increasingly lower number of particles as \psad~increases. Since for a fixed subhalo mass, the subhalo mass fraction decreases with halo mass \citep[e.g.][]{Gao_11}, the dwarf-size simulation has the poorest sampling despite its better mass (and spatial) resolution (see Table~\ref{table_sims}). 

\begin{figure}
\center{
\includegraphics[height=8.5cm,width=8.5cm]{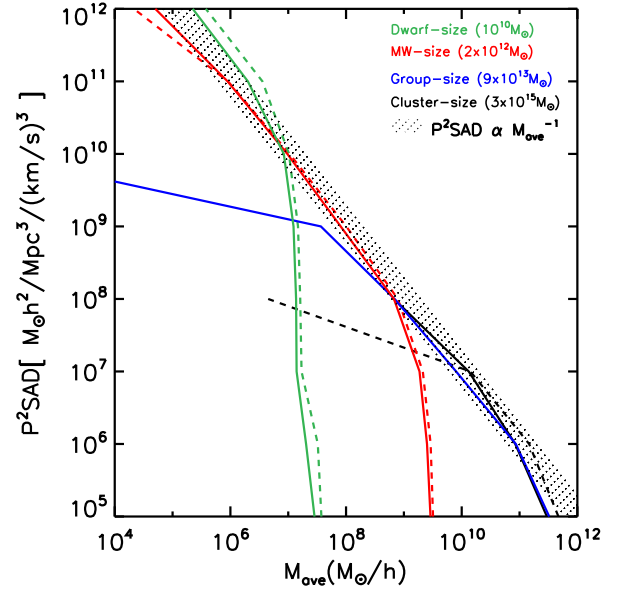} 
}
\caption{The particle phase space average density, or \psad,  in subhaloes as a function of the average mass enclosed by a \psad~contour (see Eq.~\ref{m_ave}) in four haloes of different sizes at $z=0$: cluster-size (black), group-size (blue), Milky-Way-size (red) and dwarf-size (green).
Two resolution levels are shown: high (solid) and low (dashed).} 
\label{fig_p2sad_mass} 
\end{figure}

Fig.~\ref{fig_p2sad_mass} demonstrates a different way to see the universality of \psad~and its natural cutoffs on large and small scales for a given host. It shows \psad~as a function of the average mass enclosed by each \psad~contour:
\begin{equation}\label{m_ave}
	M_{\rm ave}(\Xi_*)=\int_{\rm A(\Xi_*)} \Xi(\Delta x', \Delta v')_{{\cal V}_6} d^3{\bf \Delta x'}d^3{\bf \Delta v'},
\end{equation} 
where $A(\Xi_*)$ is the area in ($\Delta x,\Delta v$) defined by a fixed value of $\Xi_*$, i.e., the area under a given \psad ~$= \Xi_*$ contour. Each halo has two effective cutoffs in this plot: one to the right, given by the maximum mass (size) of its subhaloes, and one to the left given by the minimum mass of subhaloes that can be resolved. In the regions that are not affected by these cutoffs, to a very good approximation, $\Xi\propto M_{\rm ave}^{-1}$, which is a prediction of the spherical collapse model in a $\Lambda$CDM cosmology at small scales \citep{Afshordi_10}.

\subsection{Modelling of \psad}

In Papers I and II, we used two models of \psad, one is simply a fitting formula where the contours of $\Xi(\Delta x, \Delta v)$ are described by a family of superellipses with axes that are functions of $\Xi$ (see Eqs. 2 and 3 of Paper II):
\begin{equation}\label{lame}
\left(\frac{\Delta x}{\mathcal{X}(\Xi)}\right)^{\beta} + \left(\frac{\Delta v}{\mathcal{V}(\Xi)}\right)^{\beta} = 1,
\end{equation}
Previously, the functions $\mathcal{X}$ and $\mathcal{V}$ were modelled as power laws, but we have found that a better fit to the simulation data at all scales is given by the following functional form:
\begin{equation}\label{axes_lame}
\mathcal{X}(\Xi)={\rm atan}\left(\frac{{\rm log(\Xi)}}{p_2}\right)p_0\Xi^{p_1},
\mathcal{V}(\Xi)={\rm atan}\left(\frac{{\rm log(\Xi)}}{p_5}\right)p_3\Xi^{p_4}.
\end{equation}
Notice that for log($\Xi$)~$\gg p_2$($p_5$), these functions approach power laws, which are good approximations to the structure of \psad~at small scales
(see also Fig. 1 of Paper II).

\begin{table*}
\centering
%\resizebox{14.0cm}{!}{
\begin{tabular}{ccccccccc}
\hline
Parametric model (Eq.~\ref{lame}) & $p_0$[Mpc/h] & $p_1$ & $p_2$ & $p_3$[km/s] & $p_4$ & $p_5$ & & $\beta$ \\
\hline
\hline
 & $73.31$ & $-0.40$ & $31.04$ & $3.69\times10^4$ & $-0.30$ & $7.00$ & &1.0 \\
\hline
Physical model (Eq.~\ref{full})    &   $A_{\rm tid}$  & $\alpha$ &  $f$ & $\tilde{B}$ & $\kappa$ & $a$ & $b$ & $\beta$ \\  
\hline
\hline
  &   0.12  & 1/3 &  1.5 & 0.192 & 2.5 & 0.75 & 3.53 & 1.0 \\  
\hline
\end{tabular}
%}
\caption{Best fit parameters of two different models that describe \psad: parametric (Eq.~\ref{lame}) and physical (Eq.~\ref{full}). We note that the parameter $\beta$ is always fixed to 1.0, so these models have effectively 6 and 7 free parameters, respectively.}
\label{table_params} 
\end{table*}

The second model is a physically motivated approach that combines the stable clustering hypothesis in phase space \citep{Afshordi_10}, the spherical collapse model, and tidal disruption of subhaloes (see Sec. 4 of Paper II). In this model, structures form according to the spherical collapse model with a characteristic mass, $m_{\rm col}$, and phase space density, $\xi_s(m_{\rm col})$, at the time of collapse: 
\begin{equation}\label{rho_ps}
\xi_s=\frac{\rho_{\rm char}}{\sigma_{\rm vir}^3}=\frac{10H(\xi_s)}{G^2m_{\rm col}(\xi_s)},
\end{equation}
where the characteristic density and velocity of the collapsed object are \citep[e.g.][]{Afshordi_02}: $\rho_{\rm char}\equiv200\rho_{\rm crit}$ and $\sigma_{\rm char}\equiv\sigma_{\rm vir}=10Hr_{\rm 200}$. The subhalo collapses when the r.m.s top-hat linear overdensity $\sigma(m_{\rm col})$ (mass variance)
crosses the linear overdensity threshold $\delta_c\sim1.686$ at an epoch given by:
\begin{equation}\label{hubble}
H(\xi_s)\sim H_0\left(\frac{\sigma(m_{\rm col})}{\delta_c}\right)^{3/2}.
\end{equation}

The primordial phase space densities of structures are eventually diluted in time due to tidal stripping by a fraction $\mu(m_{\rm col})$:
\begin{equation}\label{connect}
\Xi(\Delta x, \Delta v)=\mu(m_{\rm col})\xi_s(m_{\rm col}),
\end{equation}
which is fully determined by a tidal stripping model that has 5 free parameters: the normalization ($A_{\rm tid}$) and slope ($\alpha$) of the power law dependence of tidal stripping as a function of the ambient density, a characteristic host mass where stripping begins to be effective ($fm_{\rm col}$), and the normalization ($\tilde{B}$) and slope ($\kappa$) that
define the initial condition (pre-infall) as a function of the mass variance $\sigma(m_{\rm col})$. 

The functional form of $\Xi(\Delta x, \Delta v)$ is given by the general solution to the collisionless Boltzmann equation under the stable clustering hypothesis \citep{Afshordi_10}; for $\Xi={\rm const.}$, we have:
\begin{equation}\label{full}
\left[\left(\frac{\Delta x}{a\lambda(m_{\rm col})}\right)^{\beta} + \left(\frac{\Delta v}{b\zeta(m_{\rm col})}\right)^{\beta}\right]_{\Xi={\rm const}} = 1, 
\end{equation}
where $\lambda$ and $\zeta$ are fully determined by the model (see Eqs. 17 and 18 of Paper II), while $a$, $b$ and $\beta$ are free parameters that represent $\mathcal{O}(1)$ deviations over our model to be calibrated to simulations.

Eqs.~\ref{lame} and \ref{full} describe the individual contours of \psad~from the simulation data very precisely if we fix $\beta=1$ and fit the 6 (7) additional parameters of the parametric (physical) model. Across the seven orders of magnitude of \psad~resolved in the simulations, we can find an average fit that reasonably describes the structure of \psad. Table~\ref{table_params} shows the best fit parameters for the models (given by the average values of the logarithmic least squares fit to each contour), while Fig.~\ref{fig_fits} compares them to the simulations.

\section{Clustering in subhaloes and dark matter signals}

\begin{figure}
\center{
\includegraphics[height=8.5cm,width=8.5cm]{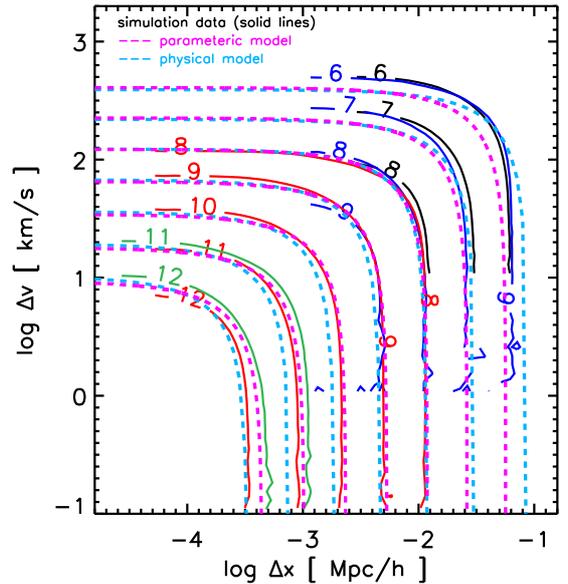} 
}
\caption{Fits to the particle phase space average density (\psad) measured in simulations (solid lines with colors, see Fig.~\ref{fig_p2sad_subs}), using a parametric formula (magenta dashed lines, Eq.~\ref{lame}) and a physically motivated model based on stable clustering + spherical collapse + tidal disruption (blue dashed lines, Eq.~\ref{full}). The best fit parameters for each model are given in Table~\ref{table_params}.}
\label{fig_fits} 
\end{figure}

From the definition of \psad, we can write the real space two point correlation function of densities as: 
\begin{eqnarray}\label{real_2pcf}
	\left<\rho({\bf x})\rho({\bf x}+{\bf\Delta x})\right>_{{\cal V}_6}&=&\int d^3{\bf v}d^3{\bf \Delta v}\left<f\right>_{{\cal V}_6}\Xi(\Delta x, \Delta v)_{{\cal V}_6}\nonumber\\
	&=&\left<\rho\right>_{{\cal V}_6}\int d^3{\bf \Delta v}~\Xi(\Delta x, \Delta v)_{{\cal V}_6},
\end{eqnarray}
where $\left<\rho\right>_{{\cal V}_6}$ is the average dark matter density within the volume where \psad~is calculated. The standard two point correlation function simply follows from Eq.~\ref{real_2pcf}:
\begin{equation}\label{real_2pcf_std}
	\xi(\Delta x)_{{\cal V}_6} = \frac{\left<\rho\right>_{{\cal V}_6}}{\bar{\rho}_m^2}\int d^3{\bf \Delta v}~\Xi(\Delta x, \Delta v)_{{\cal V}_6} - 1,
\end{equation}
where $\bar{\rho}_m$ is the mean cosmic dark matter density. The dimensionless power spectrum can then also be computed:
\begin{eqnarray}\label{power}
\Delta^2(k)&=&\frac{1}{2\pi^2}k^3P(k)_{{\cal V}_6}\nonumber\\
&=&\frac{4\pi}{2\pi^2}k^3\int_0^\infty\xi(\Delta x)_{{\cal V}_6}\frac{{\rm sin} k\Delta x}{k\Delta x}(\Delta x)^2 d\Delta x
\end{eqnarray}

To normalize $\xi(\Delta x)_{{\cal V}_6}$ to a cosmic volume $V$, we make the following substitution in Eq.~\ref{real_2pcf_std}:
\begin{equation}\label{normalization}
	\frac{\left<\rho\right>_{{\cal V}_6}}{\bar{\rho}_m^2}=\frac{1}{\bar{\rho}_m}\frac{M_{V}}{\bar{\rho}_m V}=\frac{f_{\rm subs}(V)}{\bar{\rho}_m},
\end{equation}
where $f_{\rm subs}(V)$ is the subhalo mass fraction within the cosmic volume $V$, which is approximately given by $f_{\rm subs}\sim f_{\rm subs}(V_{\rm MW})f_{\rm h}\sim0.12$, where
$f_h$ is the halo mass fraction, and $f_{\rm subs}(V_{\rm MW})$ is the subhalo mass fraction in a Milky-Way-size halo.
For  $m_{\rm min}=10^{-6}$M$_\odot$, $f_h\sim0.8$ (e.g. \citealt{Angulo_10}) and $f_{\rm subs}\sim0.15$ (e.g. \citealt{Springel_08}). 

\begin{figure}
\center{
\includegraphics[height=8.5cm,width=8.5cm]{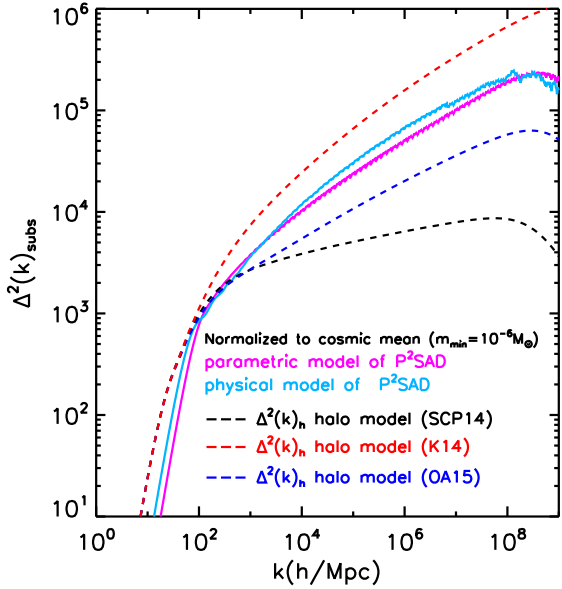} 
}
\caption{Predictions from the modelling of \psad~for the power spectrum 
of DM {\it inside subhaloes} with collapse mass in the range $10^{-6}-2\times10^{12}$~M$_\odot$, normalized to the cosmic mean density. The lower value represents the canonical 
minimum self-bound mass for 100 GeV WIMPs, while the maximum mass is given by the  minimum value of \psad~that we can trust (log(\psad)$_{\rm min}=6$, see Fig.~\ref{fig_fits}).
We also show the {\it halo model} predictions for the power spectrum in haloes in the same mass range using the mass function from \citet{Schneider_13}, and either a NFW profile with two different concentration-mass relations (SCP14, \citealt{SC_14}, dashed black, and OA15, \citealt{Okoli2015}, dashed blue), or a Einasto profile with parameters given in \citet{Klypin_14} (K14, dashed red).}
\label{2pcf_power} 
\end{figure}

In Fig.~\ref{2pcf_power} we show the dimensionless power spectrum {\it in subhaloes} as predicted by \psad. To calculate the power spectrum, we have used the {\tt FFTLOG} code \citep{Hamilton_00} to compute the Fourier transform of $\xi(\Delta x)_{{\cal V}_6}$. To make this plot we have used our models of \psad~and compute $\xi(\Delta x)_{{\cal V}_6}$ using as limits for the integral in velocities those that correspond to the values of $\Xi$. The minimum is given by the level at which we can trust the models in its comparison with simulations (log(\psad)$_{\rm min}=6$, see Fig.~\ref{fig_fits}), while the maximum is given by the CDM particle model one wishes to study. As an illustration, we have chosen a 100 GeV WIMP with a canonical value for the minimum self-gravitating mass for collapse $m_{\rm min}=10^{-6}$M$_\odot$. Since our physical model for \psad~allows us to connect a collapse mass with a value of $\Xi$ (see Eq.~\ref{connect}), we can easily interchange $m_{\rm col}$ with $\Delta v$ (for a given $\Delta x$) in Eq.~\ref{real_2pcf_std}. In this way, Fig.~\ref{2pcf_power} shows the DM clustering in subhaloes with collapse masses in the range $10^{-6}-2\times10^{12}$M$_\odot$. Remarkably, both models of \psad~give a very similar result, despite the fact that they are only calibrated within the scales resolved in the simulation. At large scales the clustering is artificially suppressed by the minimum limit we imposed on \psad, while at small scales, $\Delta^2(k)\sim{\rm const.}$, due to the physical cut-off given by $m_{\rm min}$. 

In Fig.~\ref{2pcf_power}, we also show the DM power spectrum according to the {\it halo model} \citep[e.g.][]{Seljak_00}, for halo masses in the same range as the one used for \psad:
$10^{-6}-2\times10^{12}$M$_\odot$. For the ingredients of the halo model we used the halo mass function for a Planck cosmology using the code by \citet{Schneider_13}, and two different halo profiles, NFW profile with a {\it shallow} concentration mass relation (black dashed line, \citealt{SC_14}) and Einasto profile with an effective {\it steep} concentration mass relation (red dashed line, \citealt{Klypin_14}); we also show the result for NFW profile with the concentration mass relation given by \citet{Okoli2015} (blue dashed line). It is important to remember that since \psad~was defined {\it only} with particles inside subhaloes, our results cannot be directly compared with the halo model, which is defined by the DM clustering inside (and across) haloes. At small scales however, the clustering is dominated by the so-called one-halo term, which is defined as correlations between particle pairs within collapsed haloes (subhaloes). Because of this, \psad~ predicts the dominant contribution to the small-scale  $\Delta^2(k)$, or $\xi(\Delta x)$, for {\it all} matter, albeit with a normalization that depends on the
uncertain total DM mass contained in subhaloes (see Eq.~\ref{normalization}). 

Our prediction of the non-linear power spectrum at small scales significantly decreases the uncertainty using other methods/extrapolations (see e.g., Fig. 3 of \citealt{Serpico_14} for a plot containing several of these predictions). 
In the {\it halo model} for instance, this uncertainty is ultimately connected with the sensitivity of the extrapolation (towards low-masses) of the concentration-mass relation on the assumptions regarding the structure of haloes in the resolved regime. This is illustrated in the difference between the cases SCP14 (black dashed line) and K14 (red dashed line) in Fig.~\ref{2pcf_power}. These two cases are based on calibrations with similar simulation data but using different profiles (NFW and Einasto, respectively). Their predictions for the concentration-mass relation at low masses, and therefore on the power spectrum, are quite different. 
Our method on the other hand, is based on the modelling of a 2D function (\psad), which is resolved over many orders of magnitude. 
The fact that our two different models (physical and parametric) give a very similar result for the power spectrum at small scales 
(magenta and blue solid lines in Fig.~\ref{2pcf_power}) indicates, in contrast to the {\it halo model}, a low sensitivity to the way the fit (calibration) is done 
in the resolved regime. We argue that this is an advantage of our method compared to the halo model.

In this work we have concentrated in analysing DM-only simulations. In the presence of baryons, the DM phase space clustering is expected to be modified as the DM responds dynamically to the assembly of galaxies. We anticipate that the structure of \psad~will be modified mainly at the scales where subhaloes contain condensed baryons (gas and stars). The focus of this work is however at the scales of low-mass subhaloes, which are devoid of cold condensed gas and stars. Therefore, baryons can only impact \psad~at these small scales indirectly through the orbital interaction of subhaloes with the
host galaxy and its satellites. This is mostly important for subhaloes that at any point in their orbit were near the centre of the host galaxy. Since in this work 
we average \psad~globally within the virial radius of the host halo, the average is dominated by the abundant subhaloes near the virial radius, most of which have recent infall times. Thus, at small scales, \psad~(averaged in this way) is not likely to be affected by the presence of baryons.

\subsection{Global subhalo boost to DM annihilation}

\begin{figure}
\center{
\includegraphics[height=8.5cm,width=8.5cm]{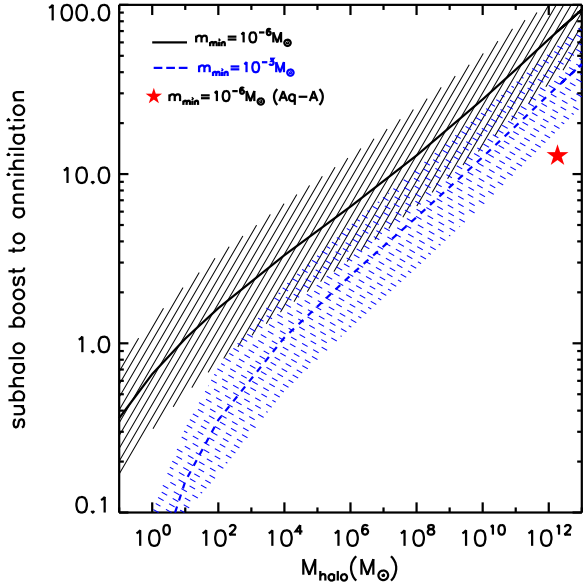} 
}
\caption{Predicted global subhalo boost (i.e. within $r_{200}$) to the annihilation rate of a host halo as a function of its mass, according to our physically motivated model of \psad~(Eq.~\ref{full}). The two cases are for two different minimum subhalo masses as shown in the legend. The annihilation rate in the smooth host halo is computed assuming it follows a NFW profile with the concentration-mass relation given in \citet{SC_14}, with a $\pm\sigma$ scatter of 0.14 dex. The star symbol is the boost for the Aquarius A halo, which has a concentration of $\sim19$.}
\label{fig_boost} 
\end{figure}

The DM annihilation rate $R_{\rm ann}$ (number of self-annihilation events per unit time) in a region of physical volume $V$ (phase space volume ${\cal V}_6$) is given by the limit of \psad~on small distances. The global subhalo boost to the annihilation rate of the smooth host halo with a density profile $\rho({\bf x})$ is (Eqs. 4, 5 and 9 of Paper II):
\begin{eqnarray}\label{rate}
  \frac{R_{\rm ann}^{\rm subs}}{R_{\rm ann}^{\rm smooth}} = \frac{M_{{\cal V}_6}\int d^3{\bf \Delta v} (\sigma_{\rm ann} v) \lim_{\Delta x \to 0} \Xi(\Delta x, \Delta v)_{{\cal V}_6}}{{\int_V d^3{\bf x}\rho^2({\bf x})\langle\sigma_{\rm ann} v\rangle}},
\end{eqnarray}
where $(\sigma_{\rm ann} v)$ is the product of the annihilation cross section and the relative velocity between pairs, and $\langle\sigma_{\rm ann} v\rangle$ is its average over the velocity distribution of DM particles.  Using the physical modelling of \psad, we can transform the numerator in Eq.~\ref{rate} into an integral over $m_{\rm col}$, which in the case $(\sigma_{\rm ann} v)={\rm const.}$, takes a simple form (Eq. 32 of Paper II):
\begin{eqnarray}\label{rate_cte_model}
  R_{\rm ann}^{\rm subs}&=&\frac{8\pi^{1/2}b^3}{9\delta_c^3}200\rho_{\rm crit,0}M_{{\cal V}_6}\frac{(\sigma_{\rm ann} v)}{2m_\chi^2}\nonumber\\
  &&\times\int_{\rm m_{min}}^{\rm m_{max}} \mu(m_{\rm col})m_{\rm col}^{-2} d(m_{\rm col}^2\sigma^3(m_{\rm col})),
\end{eqnarray}	
where  $m_\chi$ is the mass of the dark matter particles, $\rho_{\rm crit}$ is the critical density, and $m_{\rm min (max)}$ is the minimum (maximum) collapse mass for the subhaloes contributing to \psad.

Using the parameters in Table \ref{table_params}, we can use Eqs.~\ref{rate} and \ref{rate_cte_model} to compute the global subhalo boost to the annihilation rate of host haloes of any mass (Fig.~\ref{fig_boost}). To compute the total mass within subhaloes, $M_{{\cal V}_6}$, we use the formula given in \citet{Gao_11}. We show two values of $m_{\rm min}$, 
and assume a NFW profile for the host halo with a concentration-mass relation as given \citet{SC_14}. Notice that the boost strongly depends on the concentration of the smooth host halo, e.g., if we take the Aq-A simulation, which has a very high concentration for its mass, we get a lower value for the boost (red star symbol), consistent with our previous results (see Fig. 5 of Paper II). 

Our predictions for the subhalo boost are fairly high compared to, for example, those made by \citet{SC_14} (see their Fig. 2). The reason 
for this is apparent in Fig.~\ref{2pcf_power}, where we show the wide range of predictions for the power spectrum at small scales (and thus indirectly the subhalo boost) according to the {\it halo model}. This uncertainty is ultimately connected with the uncertainty on the concentration-mass relation at small masses. The work by \citet{SC_14} is in the low end of these predictions and thus, predicts a lower subhalo boost than our model. It is important to mention that the latter work assumes that the concentration of a subhalo is the same as that of an isolated halo of the same mass, while simulations have shown that subhaloes are typically more concentrated. For instance, in a Milky-Way-size halo, subhaloes are on average $\sim30\%$ more concentrated 
 \citep[see e.g. Fig. 28 of][]{Springel_08}. Given the strong dependence of the annihilation rate on the subhalo concentration, the subhalo boost is expected to be a factor of $2-3$ higher than the estimate given by \citet{SC_14} once this effect is taken into account \citep[see e.g.][for a recent detailed calculation of this effect]{Bartels2015}. This would bring their prediction much closer to ours.

\section{Conclusions}

The particle phase space average density (\psad) is a coarse-grained phase space density that contains a wealth of information on the clustering of matter (Papers I and II). 
By studying DM structures with this novel statistics, we have found a new universality in the gravitational clustering of DM within subhaloes:
\begin{itemize}
\item The structure of \psad~averaged over particles inside subhaloes is universal across haloes in a wide range of masses covering
3 orders of magnitude in physical and velocity separations, from the subhaloes of dwarf-size haloes to those of cluster-size haloes (see Fig.~\ref{fig_p2sad_subs}).
\item The functional form of \psad~(averaged in subhaloes) can be described by a family of superellipse contours in the $(\Delta x, \Delta v)$ plane with axes that can be 
modelled by simple parametric formulae (see Eq.~\ref{axes_lame}), or by a physically motivated model based on the stable clustering hypothesis in phase space, the
spherical collapse model, and tidal disruption of subhaloes (see Eq.~\ref{full}).
\end{itemize}  
We have exploited this new universality of DM clustering to accurately predict the highly non-linear DM power spectrum all the way down to the decoupling limit of CDM particles (Fig. \ref{2pcf_power}). This prediction can be used
to study several potential DM signals, e.g., the subhalo boost to the annihilation of DM for different host halo masses. A code to compute \psad~using our
physical model is available upon request.

\section*{Acknowledgments}

We thank the members of the Virgo consortium for access
to the Aquarius simulation suite and our special thanks
to Volker Springel for providing access to SUBFIND and
GADGET-3.  The Dark Cosmology Centre is funded by the DNRF. JZ is supported by
the EU under a Marie Curie International
Incoming Fellowship, contract PIIF-GA-2013-62772. NA is supported by the University of Waterloo 
and the Perimeter Institute for Theoretical Physics. Research at Perimeter Institute is supported by 
the Government of Canada through Industry Canada and by
the Province of Ontario through the Ministry of Research \& Innovation. 

\bibliography{lit}

% Don't change these lines
\bsp	% typesetting comment
\label{lastpage}
\end{document}